\documentclass[preprint,superscriptaddress,prl,showpacs]{revtex4}
\usepackage{amssymb,amsmath,graphicx,amscd,xcolor,dsfont}

\begin{document}

\title{Classical enhancement of quantum vacuum fluctuations}

\author{V. A. \surname{De Lorenci}}
\email{delorenci@unifei.edu.br}
\affiliation{Instituto de F\'{\i}sica e Qu\'{\i}mica,    Universidade Federal de Itajub\'a, \\
Itajub\'a, Minas Gerais 37500-903, Brazil}
\author{L. H. \surname{Ford}}
\email{ford@cosmos.phy.tufts.edu}
\affiliation{Institute of Cosmology, Department of Physics and Astronomy, \\
Tufts University, Medford, Massachusetts 02155, USA}

\begin{abstract}
We propose a mechanism for the enhancement of vacuum fluctuations by means of a classical field. The basic idea is
that if an observable quantity depends quadratically upon a quantum field, such as the electric field, then the
application of a classical field produces a cross term between the classical and quantum fields. This cross term
may be significantly larger than the purely quantum part, but also undergoes fluctuations driven by the quantum field.
We illustrate this effect in a model for lightcone fluctuations involving pulses in a nonlinear dielectric. Vacuum
electric field fluctuations produce fluctuations in the speed of a probe pulse, and form an analog model for quantum
gravity effects. If the material has a nonzero third-order susceptibility, then the fractional light speed fluctuations
are proportional to the square of the fluctuating electric field. Hence the application of a classical electric field
can enhance the speed fluctuations. We give an example where this enhancement can be an increase of one order
of magnitude, increasing the possibility of observing the effect.
\end{abstract}

		
\maketitle

Quantum field fluctuations, such as vacuum fluctuations of the electromagnetic field, are responsible for a variety of physical
effects, including the Lamb shift and the Casimir effect. In some cases, the effect can be expressed in terms of a time
average of a field operator or of a square of the operator. For example, the one-loop QED vertex correction to quantum potential
scattering can be interpreted as due to fluctuations of the averaged electric field~\cite{HF15}. Other effects, such as those
associated with the stress tensor, are quadratic in the fields. Quantum fluctuations of the gravitational field can be of either 
variety, and may lead to lightcone fluctuations~\cite{F95}.
The active fluctuations of spacetime geometry due the quantum nature of gravity itself can be linear, whereas the passive 
fluctuations driven by quantum stress tensor fluctuations are quadratic in the matter fields. Both types of lightcone fluctuations
can be modeled by nonlinear optical materials, where fluctuations of either the electric field or the squared electric field can
produce fluctuations in the speed of a probe pulse~\cite{FDMS13,BDF14,BDFS15,BDFR16}.

In systems where the observable effect depends nonlinearly upon the field, it may be possible to enhance the fluctuations
by the application of a classical field. Consider the square of the electric field, $E^2$, and suppose that 
$\mathbf{E} = \mathbf{E}_C + \mathbf{E}_Q$, where $\mathbf{E}_C$ is a classical electric field, and  $\mathbf{E}_Q$
is the fluctuating quantum field. Then $E^2 = E_C^2 +2 \mathbf{E}_C \cdot \mathbf{E}_Q +  E_Q^2$. The square of the classical
field does not fluctuate, and $E_Q^2$ describes the effect in the absence of the classical field. If we can arrange that
$2 |\mathbf{E}_C \cdot \mathbf{E}_Q| >  E_Q^2$, then the quantum fluctuation effects can be enhanced by the presence of the
classical field. We can make this statement more precise by relating the observable quantities to expectation values of
time averages of the quantum fields. Note that in the vacuum state, the expectation values of the quantum field, and hence of the
cross term  vanish, $\langle \mathbf{E}_Q \rangle = \langle \mathbf{E}_C \cdot \mathbf{E}_Q \rangle =0$. However, the classical
field does give a nonzero contribution to the variance of $E^2$.
Here we treat the explicit example of the nonlinear optics model for lightcone fluctuations.

In a nonlinear material, the presence of a background field, $\mathbf{E}^0$, can alter the effective index of refraction and hence
the speed of propagation of a probe pulse through the material.  The change in the effective index of refraction can be linear in the
background field (Pockels effect), or quadratic in this field (Kerr effect). If the background field fluctuates, then the propagation
speed will also be subject to fluctuations. The optical properties of a material are described by the various susceptibilities
which appear in the induced polarization, or dipole moment density. 
Unless stated otherwise, we  use Lorentz-Heaviside units with $\hbar=c=1$. The conversion between these units and SI units
can be facilitated by noting that in our units, $\epsilon_0 \approx 8.85 \times 10^{-12}\, {\rm C^2/(N \,m^2)} =1$,  implying that $1V \approx 1.67\times 10^7 {\rm m}^{-1}$.

The induced polarization vector $\mathbf{P}$ of a nonlinear optical material, can be expanded as a power series in the electric field as
$P_i = \chi_{ij}^{{}_{(1)}} E_{j} + \chi_{ijk}^{{}_{(2)}} E_{j}E_{k} + \chi_{ijkl}^{{}_{(3)}} E_{j}E_{k}E_{l} + \cdots  \,$,
where $\chi_{ij}^{{}_{(1)}}$ are the components of the linear susceptibility tensor, while  $\chi_{ijk}^{{}_{(2)}}$ and 
$\chi_{ijkl}^{{}_{(3)}}$,  are the second- and third-order nonlinear optical susceptibilities of the medium~\cite{boyd2008}, respectively. 
We use the convention that repeated indices are summed upon. The susceptibilities are generally dependent on frequency. 
However most  materials exhibit approximately constant susceptibilities within a certain range of frequencies, 
defining a dispersionless regime, which will be assumed here. We follow the procedure in 
Refs.~\cite{FDMS13,BDF14,BDFS15,BDFR16}, in which the electric field is written as a superposition of a background
field $\mathbf{E}^0$ and a smaller, but more rapidly varying probe field $\mathbf{E}^1$. To leading order, the probe
field satisfies a linear wave equation. If we take this field to be propagating in the $x$-direction, but have linear polarization
in the $z$-direction, $\textbf{E}^1=E^1(x,t){\bf \hat{z}}$,   the  equation is
\begin{equation}
\frac{\partial^2 E^1}{\partial x^2} - \frac{1}{v_{ph}^2} \frac{\partial^2 E^{1}}{\partial t^2}  = 0\,.
\label{eq:we1}
\end{equation}
Here 
\begin{equation}
v_{ph}^2=\frac{1}{{n_p}^2}\left[1+2\gamma_i E_i^{0}+3\gamma_{ij} E_i^{0}E_j^{0}\right]^{-1}\,,
\label{eq:vph}
\end{equation}
where  
\begin{equation}
\gamma_i = \frac{\chi^{{}_{(2)}}_{z(zi)}}{{n_p}^2}\,,
\label{e2a}\\
\end{equation}
and
\begin{equation}
\gamma_{ij} = \frac{1}{{n_p}^2}\left(\frac{\chi^{{}_{(3)}}_{zzij}+\chi^{{}_{(3)}}_{zizj}+\chi^{{}_{(3)}}_{zijz}}{3}\right)\,,
\label{e2}
\end{equation}
with  $\chi^{{}_{(2)}}_{z(zi)} = (\chi^{{}_{(2)}}_{zzi} +\chi^{{}_{(2)}}_{ziz})/2$. That is, the parentheses denote symmetrization
on the pair of enclosed indices. 

In a dispersionless regime, the phase velocity $v_{ph}$  is also approximately the group velocity of wave packets,
and the flight time is proportional to $\int dt/v_{ph}$. Following Refs.~\cite{BDFS15,BDFR16} we introduce a sampling
function $F(x)$ which describes the density profile of a slab of material and also acts as a switching function for the 
electric field fluctuations. It has the normalization
\begin{equation}
\int_{-\infty}^\infty F(x) \, dx = d\,,
\label{eq:norm}
\end{equation}
where $d$ is the effective width of the slab.
The flight time operator is given by
\begin{align}
t_d = n_p\int_{-\infty}^\infty \left[ {1}+\gamma_iE_i^{{}_0}(\textbf{x},t)+\mu_{ij} :E^{{}_0}_i(\textbf{x},t)E^{{}_0}_j(\textbf{x},t):
\right]F(x)\,dx \,,
\label{td}
\end{align} 
where an expansion to second order in $E^{{}_0}$ has been performed, and we  have defined
\begin{equation}
\mu_{ij} = \frac{1}{2}\left(3\gamma_{(ij)}-\gamma_i\gamma_j\right)\,.
\label{mu}\\
\end{equation}
Thus the probe pulse flight time can depend nonlinearly upon the background field.

In Refs.~\cite{BDFS15,BDFR16}, $\mathbf{E}^0$ was taken to be the quantized electric field operator, and the
state to be the vacuum, so the background field arises from vacuum fluctuations. Now we wish to add a classical
electric field  $\mathbf{E}^C$ and write
\begin{equation}
\mathbf{E}^0 = \mathbf{E}^C + \mathbf{E}^Q\,,
\end{equation}
where now $\mathbf{E}^Q$ is the quantized electric field operator

The fight time operator can be written as $t_d =  t^{{}_C}_d  + t^{{}_Q}_d + t^{{}_{Cross}}_d$, where
\begin{eqnarray}
t^{{}_C}_d &=& n_p \int_0^d \left(1+\gamma_i E^{{}_C}_i +\mu_{ij} E^{{}_C}_i E^{{}_C}_j \right)F(x)dx,
\label{clas}
\\
t^{{}_Q}_d &=& n_p \int_0^d \left(\gamma_i E^{{}_Q}_i +\mu_{ij} :E^{{}_Q}_i E^{{}_Q}_j: \right)F(x)dx,
\label{quant}
\\
t^{{}_{Cross}}_d &=& 2 n_p \int_0^d \mu_{ij}  E^{{}_C}_i E^{{}_Q}_j F(x)dx.
\label{mix}
\end{eqnarray}
Note that $t^{{}_{Cross}}_d$ is a cross term coupling classical and quantum contributions of the background electric field.
 
The vacuum expectation value of $t_d$ is just $t^{{}_C}_d$, as $\langle t^{{}_Q}_d\rangle =0$ and $\langle t^{{}_{Cross}}_d\rangle =0$. 
However, as $t^{{}_C}_d$ is a c-number, only  $t^{{}_Q}_d$ and $t^{{}_{Cross}}_d$ contribute to the variance of the flight time, i.e., 
\begin{align}
(\Delta t_d)^2 = & \;\langle t_d{}^2\rangle - \langle t_d\rangle^2 = \langle( t^{{}_Q}_d + t^{{}_{Cross}}_d)^2\rangle=
\nonumber\\
=&\; {n_p}^2\int_{-\infty}^{\infty}dx \,F(x)\int_{-\infty}^{\infty}dx' \,F(x') \Big[\Gamma_i  \, \Gamma_j \,
\langle E_i^{{}_Q}(\textbf{x},t)E_j^{{}_Q}(\textbf{x}',t')\rangle
\nonumber\\
&+\mu_{ij}\mu_{lm}\langle:E_i^{{}_Q}(\textbf{x},t)E_j^{{}_Q}(\textbf{x},t)::E_l^{{}_Q}(\textbf{x}',t')E_m^{{}_Q}(\textbf{x}',t'):\rangle\Big].
\label{dispersion}
\end{align}
where we have defined
\begin{equation}
\Gamma_i  = \gamma_i + 2\mu_{ki} E_k^{{}_C}\,.
\end{equation}

Next, we use Wick's theorem to simplify the last term in Eq. (\ref{dispersion}), and introduce the needed correlation functions 
of the electric field for a non-dispersive isotropic material with refractive index $n_b$~\cite{BDFR16}. We then obtain
\begin{align}
(\Delta t_d)^2 =	\int_{-\infty}^{\infty}dx \,F(x)\int_{-\infty}^{\infty}dx'F(x')\left[\frac{A}{(\Delta x)^4}+\frac{\alpha }{(\Delta x)^8}\right],
\label{final2}
\end{align}
where we have defined the $A$ and $\alpha$ parameters by
\begin{align}
A =& \frac{n_b{n_p}^2}{\pi^2\left({n_p}^2-{n_b}^2\right)^2}\left[\Gamma_x^2 + \left(\Gamma_y^2+ \Gamma_z^2\right)\frac{\left({n_p}^2
+{n_b}^2\right)}{\left({n_p}^2-{n_b}^2\right)}\right],
\label{a1}\\
\alpha =& \frac{2{n_b}^2{n_p}^2}{\pi^4\left({n_p}^2-{n_b}^2\right)^4}\left[2\left(\mu_{xy}^2+\mu_{xz}^2\right)
\frac{\left({n_p}^2+{n_b}^2\right)}{\left({n_p}^2-{n_b}^2\right)}\right.
\nonumber\\
&\left.+\mu_{xx}^2 + \left(\mu_{yy}^2+\mu_{zz}^2+2\mu_{zy}^2\right)\frac{\left({n_p}^2+{n_b}^2\right)^2}
{\left({n_p}^2-{n_b}^2\right)^2}\right].
\label{a2}
\end{align}

Now, in order to describe the physical transition experienced by the probe pulse as it enters and exits the optical material, we 
choose the two parameter switching function~\cite{BDFR16} $F(x) =  F_{b,d}(x)$, where
\begin{equation}
 F_{b,d}(x) =  (1/\pi)[\arctan(x/b) + \arctan((d-x)/b)]\,,
\label{fbd}
\end{equation}
which satisfies Eq.~(\ref{eq:norm}).
Here, parameter $d$ describes the width of $F_{b,d}(x)$, while $b$ determine how fast the function rises and falls. 
For instance when $b\rightarrow 0$ we recover a step function, which describes a sudden transition between the different regimes,
and for $b\approx 0.9 d$ we find a broad function similar to a  Lorentzian. We are especially interested in the case where 
$0<b \ll d$, so the transition occurs over a finite region smaller than the width.
The derivative of $F_{b,d}(x)$ with respect to $x$ is a sum of two Lorentzian functions.
Assuming that $\Delta x = x-x' - i \varepsilon$, with $\varepsilon > 0$, the integrals in Eq. (\ref{final2}) can be evaluated by means of residue theorem, which leads to 
\begin{align}
(\Delta t_d)^2 = \frac{d^2(d^2+12b^2)}{12b^2(d^2+4b^2)^2} \,  A
+   \frac{d^2(21504b^{10}+1344b^6d^4+240b^4d^6+24b^2d^8+d^{10})}{1344b^6(4b^2+d^2)^6} \, \alpha,
\label{final3}
\end{align}
where we assumed the classical field to be a constant. We now define the fractional fluctuations in the flight time,
$\delta$, by
\begin{equation}
\delta^2 = \frac{ (\Delta t_d)^2}{\langle t_d\rangle^2}\,.
\end{equation}
We will be interested in the regime of $b/d <<1$, for which 
\begin{align}
\delta^2  \approx \frac{A}{12{n_p}^2d^4}\left(\frac{d}{b}\right)^2+\frac{\alpha }{1344{n_p}^2d^8}\left(\frac{d}{b}\right)^6\,.
\label{frac}
\end{align}

In the case of a centrosymmetric material, the second order coefficients $\chi^{(2)}_{ijk}$ are all  zero. Then $\gamma_i=0$, 
which leads to  $\mu_{ij} = (3/2)\gamma_{(ij)}$ and $\Gamma_i = 3\gamma_{(ij)} E_j^{{}_C}$. In this case, the first term
in Eq.~\eqref{frac} is proportional to $ (E^{{}_C})^2$, and describes the effect of the classical electric field on the
enhanced vacuum fluctuations. The second term is independent of the classical field, and describes the vacuum flucuations
in the absence of the classical field.

We may estimate the magnitudes of both effects for the case of silicon. This material has a third order susceptibility 
$\chi^{(3)}_{zzzz} \approx 2.80\times 10^{-19}{\rm m^2V^{-2}}$ and a refractive index $n_b = 3.418$, both at wavelength 
$\lambda_b = 11.8 \mu {\rm m}$ ~ \cite{Salzberg57,wynne1968,Charra2}. Assume that the probe field wave packet is prepared 
so that its peak wavelength is $\lambda_p = 1.4 \mu {\rm m}$, for which $n_p = 3.484$~\cite{Primak71}. This leads to
\begin{equation}
A \approx \;5.07 \times 10^{-36} (E^{{}_C})^2 ({\rm m^4/V^4}) \,,
\label{a1b}
\end{equation}
and 
\begin{equation}
\alpha \approx  \; 2.21 \times 10^{-34} ({\rm m^4/V^4}) \,.
\label{a2b}
\end{equation}
Now Eq.~(\ref{frac}) may be expressed as
\begin{align}
\delta^2  \approx 1.24\times 10^{-17} \left(\frac{10 \mu{\rm m}}{d}\right)^4\left(\frac{d}{b}\right)^2\left(\frac{E^{{}_C}}{E_{{}_{Break}}}\right)^2 + 1.74\times 10^{-27} \left(\frac{10 \mu{\rm m}}{d}\right)^8\left(\frac{d}{b}\right)^6\,.
\label{fracdb}
\end{align}
Here $E_{{}_{Break}} \approx 3.15\times 10^7 {\rm V m^{-1}}$ is the breakdown electric field for silicon~\cite{kay1954}, and
we need the require $E^{{}_C} < E_{{}_{Break}}$. Now set  $b=0.01d$, which corresponds to the most rapid switching
which is compatible with our neglect of dispersion~\cite{BDFR16}. Then we have the estimate
\begin{align}
\delta^2  \approx 1.24\times 10^{-13} \left(\frac{10 \mu{\rm m}}{d}\right)^4\left(\frac{E^{{}_C}}{E_{{}_{Break}}}\right)^2 + 
1.74\times 10^{-15} \left(\frac{10 \mu{\rm m}}{d}\right)^8\,.
\label{fracdb2}
\end{align}
In the absence of the classical field, the vacuum effect gives $\delta \approx 4.2 \times 10^{-8}\, (10 \mu{\rm m}/d)^4$, as found 
in Ref.~\cite{BDFR16}. However, if $E^{{}_C} \agt 0.1\, E_{{}_{Break}}$, then the first term in Eq.~(\ref{fracdb2}) 
dominates, and we have
\begin{align}
\delta  \approx 3.52\times 10^{-7} \left(\frac{10 \mu{\rm m}}{d}\right)^2\left(\frac{E^{{}_C}}{E_{{}_{Break}}}\right)\,.
\label{fracdominant}
\end{align}
 Thus the presence of the classical electric field can enhance the fractional flight time variation due to vacuum fluctuation
 by close to one order of magnitude. This may aid experimental observation of vacuum driven lightcone fluctuations.
 
 In summary, we have constructed a specific model in which application of a classical field can enhance vacuum
 fluctuations effects. This illustrates the general principle that fluctuations of a quantity which is quadratic (or higher
 power) in a quantum field can be enhanced by the presence of a classical field. This includes fluctuations of the
 stress tensor for a quantum field, which is typically quadratic in the field. Note that the probability distribution associated
 with the cross term between a quantum field and a classical field will be the Gaussian distribution for free field
 fluctuations. However,  the probability distribution associated with quantities quadratic in quantum fields can be
 very different, and sensitive to the choice of sampling function.~\cite{FFR10,FFR12,FF15}.

\vspace{0.8in}
\begin{acknowledgments}
This work was supported in part by the Brazilian research agencies CNPq (grant 302248/2015-3) and 
FAPEMIG (grant ETC-00118-15), and by the National Science Foundation (grants PHY-1506066 and PHY-1607118).
\end{acknowledgments}

\end{document}